# The Trinity High Explosive Implosion System: The Foundation for Precision Explosive Applications

Eric N. Brown* and Dan L. Borovina
Los Alamos National Laboratory
Los Alamos, NM 87545

This article is set during the 1944 and 1945 final push to complete Project Y—the Manhattan Project at Los Alamos—and focuses primarily on overcoming the challenge of creating and demonstrating a successful convergent explosive implosion to turn a subcritical quantity of plutonium into a critical mass. The critical mass would then efficiently yield kilotons of trinitrotoluene (TNT)-equivalent energy in about a microsecond, demonstrating the implosion atomic bomb concept. This work culminated in the Trinity atomic test near Alamogordo on July 16, 1945. This implosion effect demarcated the approach to explosive science and technology the Laboratory has followed ever since, including development of high-explosive synthesis and formulation, small and large test and diagnostic facilities, shock dynamics theory, high-explosive system design engineering, and three-dimensional implosion modeling and simulation using some of the fastest computers in the world. This work also ushered in a period of broader application of precision high explosives in conventional munitions, demolition, mining and oil exploration, and space travel.

## I. INTRODUCTION

The summer of 1944 at Los Alamos was a defining moment for the Manhattan Project.[1] Not only did the focus and direction of technical research and development change drastically in its aim to build a war-ending weapon by August 1945, but it also marked the beginning of a successful new era—the era of big science and technology, where technical collaboration went beyond the theoretical and developed practical applications that could make a difference in the world. Aside from ending World War II and greatly altering the direction and history of the world, the completion of the implosion project at Los Alamos, culminating in the Trinity atomic test of the "Christy Gadget"[2] near Alamogordo on July 16, 1945, proved that a multi-disciplinary, multi-facility endeavor with steadfast leadership can be successfully undertaken even though believed impossible by most, including by other major nations in the world, such as Germany, Japan, and the Soviet Union. Numerous books have been written about the broader Manhattan Project, Project Y at Los Alamos, and the Trinity test (see for example references[3-9]). Here we focus on the development of the convergent explosive implosion system that was employed in the Trinity "Gadget" test device and in the Fat Man bomb that was detonated over the Japanese city of Nagasaki on August 9, 1945.

The Manhattan Project team of dedicated scientists and engineers, led by a secret contingent in Los Alamos, NM, succeeded in their quest despite the fact that, by the spring of 1944, many important concepts needed for an implosion weapon were still practically unknown. These unknown concepts included (1) the spontaneous fission rate of reactor-made plutonium; (2) the theoretical and practical challenges of turning many diverging explosions into a single converging "implosion"; (3) the ability to do so simultaneously and in a short period of time; (4) the lack of any significant quantity of plutonium to test, and (5) the theory and safety of dynamically-critical plutonium at different densities. The challenges faced at the time by the Los Alamos group spanned many engineering technical fields as well as scientific disciplines.

The clear direction and ultimate goal of the Manhattan Project, coming from the highest national political authority, was to deliver a usable weapon so powerful that it would end the war and potentially limit armed escalations in the future. It was the original intent of Los Alamos leadership and relevant national technical committees, including the National Defense

---
*corresponding author: en_brown@lanl.gov

Research Committee (NDRC) which later became the Office of Scientific Research and Development (OSRD), that this would be accomplished through a gun weapon using either uranium or plutonium as the fissionable material. However, it became clear in the spring and summer of 1944, that the preferred gun design that later became Little Boy would not generate a technically viable path to a plutonium-based weapon. Unlike uranium 235, the surprising faster spontaneous fission rate seen in plutonium received from the B reactor at the Hanford site[10] made it practically impossible to fire a conventional gun weapon without seeing a predetonation of the nuclear material. At that point in 1944, the two options discussed for a plutonium weapon involved either a much larger and much faster (by orders of magnitude) gun weapon option or a three-dimensional implosion option, involving the dynamic compression of a subcritical quantity of plutonium into a critical mass in microseconds. Richard Tolman, chief technical adviser to General Leslie Groves on the Manhattan Project and who had received his Ph.D. in physical chemistry from the Massachusetts Institute of Technology, co-developed the Tolman–Oppenheimer–Volkoff equation[11] while on the faculty of the California Institute of Technology, and served as Chairman of the Armor and Ordnance Division of the NDRC. Tolman had proposed implosion as a mechanism to assemble a critical mass as early as 1942, but it received little interest among Laboratory scientists.

As physicists at Los Alamos struggled making implosion a reality, Oppenheimer reached out to James Conant—then chairman of the NDRC— who suggested George Kistiakowsky, a chemist, for help. Kistiakowsky was considered the "number one civilian explosives expert" at the time, and had experience with dynamic experiment diagnostics. John von Neumann, Luis Alvarez, and others eventually joined the effort and ultimately, after hundreds of experiments and months of testing, the Explosives (X) Division finally produced a promising configuration for implosion.

## II. THE OFFICE OF SCIENTIFIC RESEARCH AND DEVELOPMENT

The NDRC was hugely consequential to United States (U.S.) war time scientific advances, including secret radar development and the Manhattan Project. The NDRC was created by Vannevar Bush "to coordinate, supervise, and conduct scientific research on the problems underlying the development, production, and use of mechanisms and devices of warfare"[12] in the U.S. on June 27, 1940. Most of the NDRC's work was done with the strictest secrecy, and it researched what would become some of the most important technology during World War II. The NDRC was superseded by the OSRD on June 28, 1941. Many of the key personnel associated with leading the high explosives and implosion research at Los Alamos either came directly from the NDRC and OSRD, from other projects funded by the organizations, or through connections and recommendations from NDRC and OSRD leadership.

## III. KEY PERSONNEL

The success of the explosives work during the Manhattan project resulted from the contributions of many people. Military personnel worked alongside civilians from the local communities and those moving in from across the country and world to work on the Manhattan project. We highlight Seth H. Neddermeyer, George B. Kistiakowsky, John von Neumann, Luis W. Alvarez, and James L. Tuck as key personnel who led the conceptual design and implantation of high explosives and implosion research at Los Alamos. They are introduced in the subsequent sections.

### III.A. Manhattan Project Women in the Explosives Workforce

While about 30 percent of the Manhattan Project work force in Los Alamos were women by October of 1944, only four—Frances Dunne, Margaret Ramsey, Lilli Hornig, and Elizabeth Boggs (introduced below)—are reported to have worked in explosives and had their stories captured by Howes and Herzenberg.[8] With the overwhelming majority of the Project Y explosives work force being men and the language of the time, the people working with raw explosives powder were nicknamed 'powder men', illustrated in Figure 1.

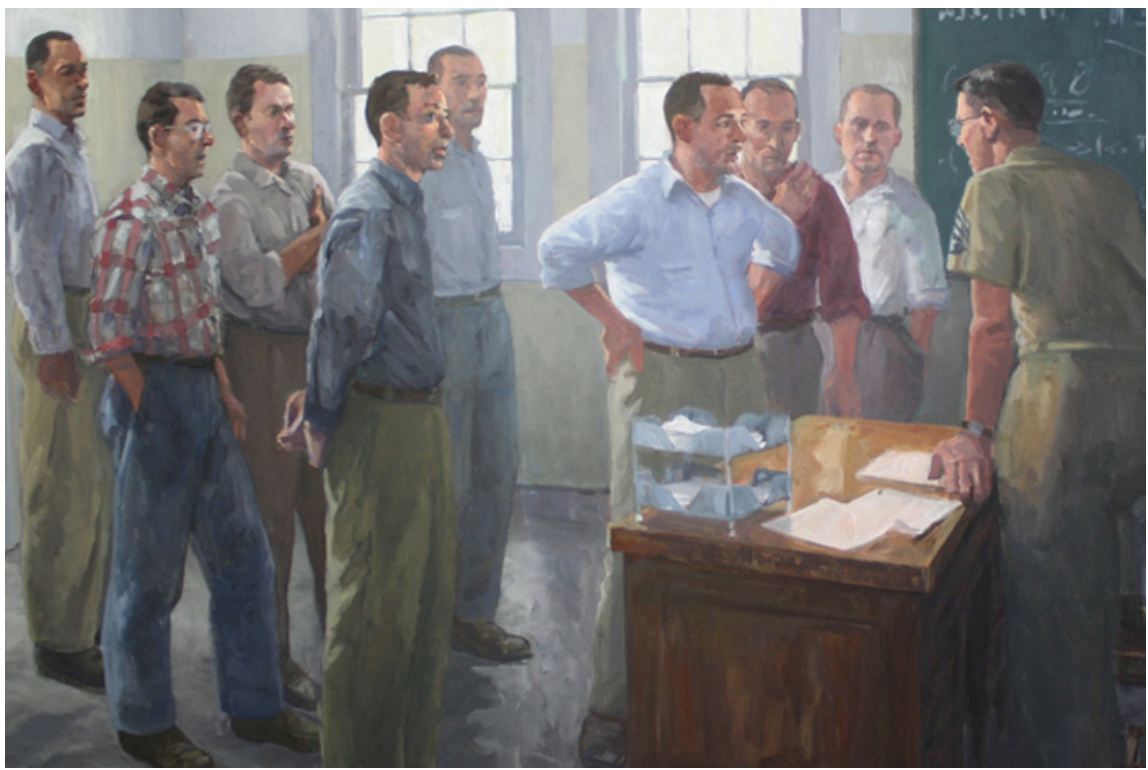

Figure 1. "The Powder Men" by John Hull depicts Los Alamos physicist McAllister Hull introducing eight newly-arrived "powder men" during the Manhattan Project years of the Laboratory.[13]

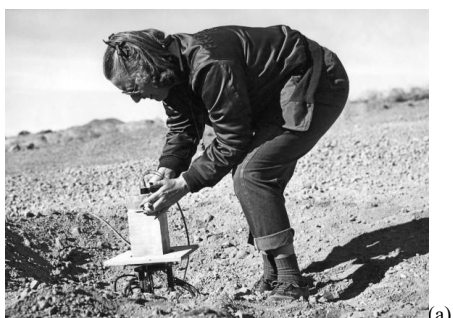
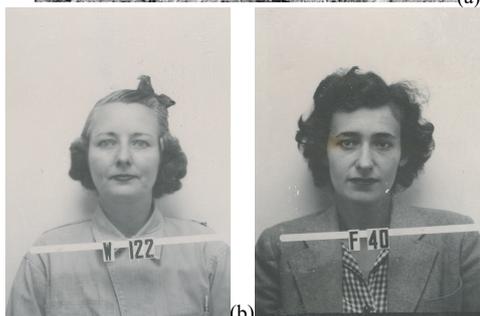

Figure 2. (a) Frances Dunne setting an explosives point in the spring of 1945 at R Site or Two Mile Mesa (Source: Los Alamos Historical Society Photo Archives). Badge pictures for (b) Frances Dunne and (c) Lilli Hornig.

Frances Dunne (Figure 2a and b), who had been a senior aircraft mechanic at Kirtland Field when the war began, was hired by Kistiakowsky into the Explosives Group for her mechanical skills. She worked as a technician and explosives supervisor in Explosives Assembly alongside thirteen tech sergeants in Los Alamos. She was the only one not to go to Tinian for Project Alberta—the wartime delivery of the completed atomic bomb—because of the United States barring women from combat. After the war she worked for the Federal Bureau of Investigation FBI).

Margaret Ramsey (1923–2016) came to Los Alamos after receiving her undergraduate degree in physics from the University of Rochester, where she worked on astrophysics. At Los Alamos she was interested in pursuing theoretical work and was assigned to the Explosives Group with a focus on the explosives used in detonators. After the war she worked on beta-decay experiments before leaving Los Alamos and eventually leaving the field of physics.

Lilli Hornig (1921–2017, Figure 2c) was born in Czechoslovakia and immigrated to the U.S. in 1933 to escape the rise of fascism. She came to Los Alamos with an almost complete Ph.D. in chemistry from Harvard, which she completed after the war. While she initially worked on plutonium purification, she transferred to the Explosives Division, where she led a section working on the development and fabrication of the explosive lenses and x-ray imaging of implosion experiments. A witness to the Trinity test, she recalls the vivid colors of the blast. Hornig went on to be a chemistry professor at Brown University, a chairwoman of the chemistry department at Trinity College in

Washington, D.C., to serve on several committees on equal opportunities including for the National Science Foundation and the American Association for the Advancement of Science, and to write several books on the topic, including "Equal Rites, Unequal Outcomes: Women in American Research Universities."[14]

Elizabeth Boggs (1913-1996) worked at the Explosives Research Laboratory in Bruceton, PA after earning her Ph.D. in Theoretical Chemistry at the University of Cambridge. Boggs had earlier demonstrated a similar independently developed explosive lens system that she showed to achieve some convergence, the details of which Duncan MacDougal—who also worked at the Explosives Research Laboratory and worked at Los Alamos for much of his later career—had sent to Los Alamos.[7] Soon after, she was recruited to work on war-related explosives research at Los Alamos. Following her son having a neonatal illness leaving him with a lifelong disability, Boggs eventually shifted her focus to advocacy and the development of public policy for people with disabilities and was a founder of the National Association for Retarded Children.

Both the field of explosives science, engineering and technology and Los Alamos National Laboratory have since become more inclusive and continue to make progress in engaging a diverse work force in contributing to national security. A rich cross section of our explosive's workforce is highlighted in our recent High Explosives Research & Development Strategy.[15] We also are benefiting from rising diversity in our highly recognized researchers, leaders, and managers contributing to field of explosives, of which we highlight just a few. Dr. Dana Dattelbaum is a senior scientist working in shock and detonation physics and recipient of the DOE E.O. Lawrence Award, an APS Fellow, a LANL Fellow, and manages of the Dynamic Material Properties Program. Dr. David Chavez is a senior scientist working in explosive synthesis and a recipient of the DOE E.O. Lawrence Award, an AAAS Fellow, and is Deputy Group Leader for the High Explosives Science and Technology group. Dr. Laura Smilowitz is a senior scientist working on imaging of the explosive deflagration and an APS Fellow, an AAAS Fellow, and a LANL Fellow. Dr. Michelle Espy is a senior scientist working on explosive detection and an APS Fellow. Dr. Margo Greenfield is the Group Leader for the High Explosives Science and Technology group. Dr. Kimberley Scott and Dr. Jennifer Young are the LANL Program Directors for the Experimental Science Program and Directed Stockpile Work Program Integration respectively, which in combination support the majority of high explosives work at the laboratory. Thomas Sisneros is the Deputy Facilities Operations Director overseeing Explosives Operations.

### III.B. Seth H. Neddermeyer

Seth Neddermeyer (1907–1958, Figure 3a)—who had received his Ph.D. in physics at the California Institute of Technology and done work at the Carnegie Institution of Washington and the National Bureau of Standards on the proximity fuze, another NDRC project[4]—was recruited to work on the Manhattan Project by Robert Oppenheimer. Seth was a strong early advocate for employing implosion. Although Oppenheimer was personally doubtful of the feasibility of the implosion method in time for the delivery deadline (set for July of 1945 to match the availability of the first significant quantity of reactor-produced plutonium), he appointed Neddermeyer as Group Leader for implosion experiments in the Ordnance Division after some promising technical analysis.[7] In collaboration with Hugh Bradner,—another physicist who had received his Ph.D. from the California Institute of Technology—and James L. Tuck—a physicist from the British Mission[16]—Neddermeyer conceived the early idea of explosive implosion. From his October 23, 1944 patent,[17] his simple concept only involved "positioning explosive material around the periphery of the fissile material so that portions of said explosive material are oppositely disposed relative to said fissile material, […] detonating the explosive material whereby the resulting force of the explosion by reason of its opposite disposition compresses the fissile material." However, the group struggled with implementing the idea of explosive implosion into a practical experiment. During a Laboratory reorganization in 1945 in which the first incarnation of M-Division was formed with Neddermeyer leading M-10, the Betatron Group.[5] Neddermeyer later submitted a more complete patent along with Tuck and von Neumann on the Trinity lens system.[18]

### III.C. George B. Kistiakowsky

In January of 1944 James Conant—then chairman of the NDRC—recruited George Kistiakowsky to join the Manhattan Project. Kistiakowsky (1900-1982, Figure 3b) was a Ukrainian-American physical chemistry professor at Harvard who had in 1940 been appointed Head of Section A-1 (Explosives) in Division B (Bombs, Fuels, Gases, Chemical Problems) of the NDRC. When the NDRC became the OSRD, Kistiakowsky became Head of Division B and subsequently the Head of Division 8 responsible for explosives and propellants, until being recruited into the Manhattan Project. Within six months of arriving at Los Alamos, Oppenheimer asked Kistiakowsky to assume the role of Associate Division Leader of the Explosives Division[4] for all work on implosion and demoted Neddermeyer to senior technical advisor. The resulting implosion design, went on to be used in the Trinity Test roughly one year later on July 16, 1945[19] at the United States Army Air Fouce (USAAF) Alamogordo Bombing and Gunnery Range and in the Fat Man bomb detonated over Nagasaki, Japan, on August 9, 1945.

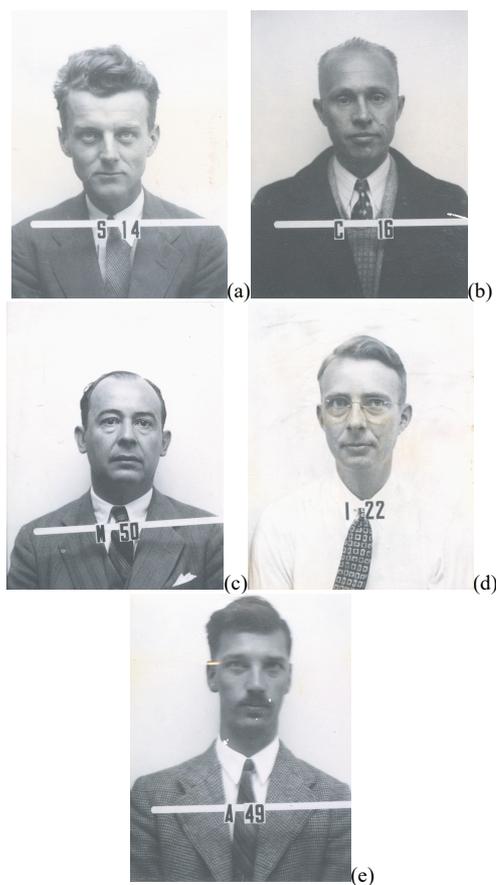

Figure 3. Badge pictures for (a) Seth H. Neddermeyer, (b) George B. Kistiakowsky, (c) John von Neumann, (d) Luis W. Alvarez and (e) James L. Tuck.

### III.D. John von Neumann

One of the Hungarian-American contributors to the Los Alamos design effort, John von Neumann (1903–1957, Figure 3c) was a mathematician and physicist, who would today be considered an extraordinary computer scientist for his modelling and simulation work. Although von Neumann made major contributions to many fields (mathematics, genetics, and others), during the Manhattan Project he worked with theoretical physicist Edward Teller, mathematician Stanislaw Ulam, and George Kistiakowsky developing the mathematical models behind the explosive lenses used in the implosion-type atomic weapon. Von Neumann also coined the term "kiloton" (of TNT-equivalent energy) as a measure of the explosive force generated by an atomic explosion, and later designed and promoted the concept of "Mutually Assured Destruction" to try and limit the subsequent nuclear arms race. Of the key personnel outlined in this paper who contributed to the development of the high-explosive implosion system for the Trinity device, he was the only one reported to be among the distinguished visitors who observed the event along with General Groves at the nearest observation point to the tower, located a mere 10,000 yards away. Everyone at this nearest observation point were instructed to lie on the ground, face downward, heads away from the direction of the blast.[4] After the war, von Neumann went to the Institute of Advanced Studies at Princeton University, but continued as a consultant in the Los Alamos Scientific Laboratory (LASL) Theoretical Division.[5]

### III.E. Luis W. Alvarez

After contributing to the secret radar project and working with Enrico Fermi at the University of Chicago, Luis Alvarez (1911–1988, Figure 3d) was recruited by Robert Oppenheimer to come to Los Alamos where he led the Electronic Detonators group, G-7, for a time and became Kistiakowsky's deputy in X (Explosives) Division. Luis was an American experimental physicist, inventor, and professor and was considered "one of the most brilliant and productive experimental physicists of the twentieth century."[20] During the Los Alamos effort, he provided a bridge wire detonator initiation system that ensured the required simultaneity of the Trinity implosion design. He also led the instrumentation team that developed and deployed a set of calibrated microphones to measure the strength of the blast wave from the atomic explosions Trinity and Little Boy. Among his many awards and acknowledgements, Alvarez was awarded the Nobel Prize in Physics in 1968 for his decisive contributions to elementary particle physics, in particular the discovery of a large number of resonance states, which was made possible through his development of the technique of using hydrogen bubble chamber and data analysis. His achievements are often celebrated as part of National Hispanic Heritage Month.

### III.F. James L. Tuck

While pursuing his Ph.D. at the Victoria University of Manchester, British-born James L. Tuck (1910–1980, Figure 3e) was pulled into the war effort in 1939 to serve as Scientific Advisor to Prime Minister Churchill's Private Office, where he was tasked with researching shaped charges that were eventually used in anti-tank weapons and for which he received the Order of the British Empire in 1944. Tuck was sent to Los Alamos in April 1944 to serve as the Principal Scientific Officer on the British Mission during the Manhattan Project.[21] His work on shaped charges, as a stepping stone into precision explosives, guided his significant contributions in designing explosive lenses for implosion.[4, 22] He outlined the lens design in his 1945 patent[23] with Hans A. Bethe as "outward of the shell a spherical arrangement of explosive lenses are incorporated to convert the nominally divergent waves created by the detonation of the multiplicity of explosive initiators to a converging spherical wave." After attending the Trinity Test and running a large diagnostics team for Operation Crossroads at Bikini Atoll, during the Laboratory reorganization in 1945 M-7, the Super Mechanics Group, was formed under Tuck's leadership to pursue experimental studies of proposed mechanical methods of initiating thermonuclear reactions.[5] He returned to the United Kingdom (U.K.) to see about completing his Ph.D. thesis but was notified by the University of Manchester that the statute of limitations on presenting a thesis had run out. While never receiving a Ph.D., he went on to serve as the Associate Division Leader for Physics Division at LANL[24] from 1956 to 1973 and was named a Fellow of the American Physical Society and American Association for the Advancement of Science.[25]

## IV. USING MULTIPLE EXPLOSIONS TO CREATE THE PERFECT IMPLOSION: "GOING AGAINST NATURE"

Developing and refining the explosives lenses for the implosion bomb was one of the greatest challenges for Los Alamos scientists and engineers. Expanding on the one-dimensional gun assembly bomb concept, the implosion concept envisioned by Seth Neddermeyer and Richard Tolman aimed to combine multiple subcritical pieces of fissionable material into a critical mass for a brief period of time, and then initiate the fission reaction using a controlled neutron source.

Conceptually this could be attained by simply increasing the external pressure of the material relative to the internal pressure until the core's density is maximized and its volume minimized. In practice, because of the time frames (microseconds) involved, this could only be attained using high explosives, and provided many challenges. Some of these challenges included using divergent high-explosive detonation shockwaves to create a single, spherically-symmetric implosion around the plutonium core, doing so simultaneously using explosive lenses with multiple surface detonators, visualizing and quantifying the detonation shockwave velocities of available (badly characterized) high explosives and combining them coherently, modeling the detonation shockwave velocities and pressures to predict the core movement and ultimate density, and designing a suite of tests using the available diagnostic capability to capture correct data and iterate design improvements.

## V. HIGH EXPLOSIVES WORK ON IMPLOSION

Whether employed for military applications or for commercial applications such as mining, explosives had always been a very crude and destructive tool. For hundreds of years before the Manhattan Project, the field of explosives, grown out of the development of gunpowder in the ninth century by Taoist Chinese alchemists, was used primarily to harness the brute divergent force created by reacting energetic chemical compounds. The Manhattan Project finally introduced and advanced the concepts of precision use, diagnostics, modeling, and control of explosives and detonation, all needed (and still employed today) for the practical realization of implosion-type fission weapons.

In an interview in 1982 with Richard Rhodes, Kistiakowsky explained the implosion concept, "Each initiation point would be centered on a lens, which would convert a divergent beam of an explosion wave into a convergent beam. Very much like an optical lens converts divergent light into a convergent light if you put it right." However, little was known at the time about how to produce imploding shock waves, or about shaping and arranging explosives to achieve controlled detonation waves. Kistiakowsky based the lens design on experiments performed by Bruno Rossi. Rossi was an Italian physicist of Jewish faith who was pressured to leave because of new Italian racial laws, and who made his way to a professorship at Cornell through multiple stops, including some time with Enrico Fermi in Chicago. He joined Los Alamos in 1943 and became the leader of the diagnostic detector group—an early example of the increasingly-critical relationship between advancements in state-of-the-art diagnostics and advancements in high explosives and weapons research and development.

The explosive lenses concept employed two explosives with different detonation velocities. When a detonator creates a point initiation in the faster explosive, a nominally divergent spherical shockwave propagates out at constant velocity. If center lit on one end of a cylinder of constant cross section, the detonation will break out at the center of the other end before doing so at its circumference, consistent with the difference in distance to the initiation point. For implosion to work, the detonation wave needs to uniformly reach the inner surface (or the "tamper") at the same time (see Figure 4). By replacing some of the faster explosive with a slower detonation velocity explosive, the breakout time at a given point on the inner surface of the explosive can be designed based on the ratio of detonation velocities of the two explosives, and on the total length between the initiation point and the surface, to determine the ratio of lengths of fast and slow explosives required along that ray.

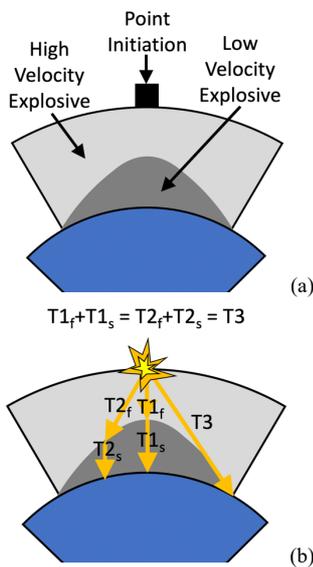

Figure 4. (a) The conceptual arrangement of a lens system in an implosion weapon. (b) Detonation shockwave travel time from the detonator to the tamper surface is the same for all paths.

The height of the cone is proportional to the detonation velocity of the slower explosive, and the slant height is proportional to the detonation velocity of the faster explosive (Figure 5). The height can be shortened relative to the base's diameter by increasing the difference between the two explosives' detonation velocities.[26-28]

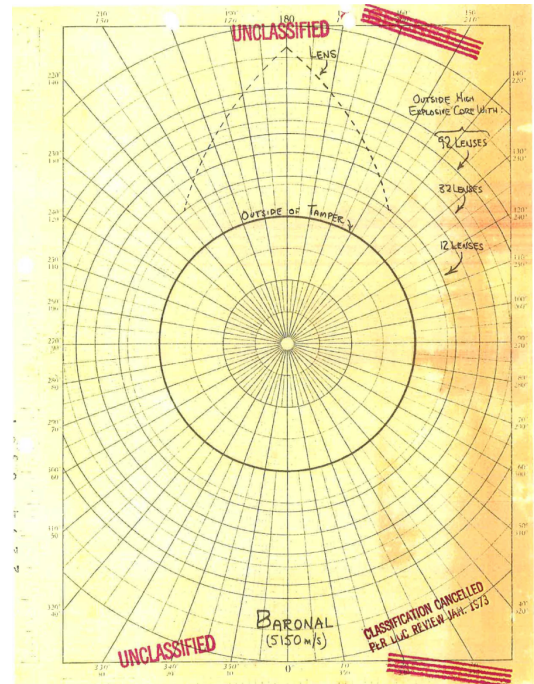

Figure 5. Unclassified lens profile design for varied numbers of lenses.[29]

This concept is still frequently used today for modern high explosive planewave lenses. Most common solid explosives exhibit detonation velocities between 3 and 10 mm/μs, although many have undesirable output pressure, sensitivity, manufacturability, stability, or mechanical properties, making them unattractive for producing explosive lenses. Kistiakowsky selected Composition B and Baratol as the two explosives for the lenses, both of which are melt castable explosives with a TNT binder (see photo in Figure 6). While the final lenses produced for the Trinity device employed Composition B and Baratol pieces that were separately melt cast and precision machined to fit, during the design George B. Kistiakowsky[29] reported that "To prosecute successfully research work on explosives lenses, several groups in the H.E. Project need a large variety of moulds and expendable components for casting and assembly of explosive lenses." To meet this demand, he went on to outline various methods for casting one explosive over the other and to manufacture the lens by packing loose explosive powder to shape for one the components.

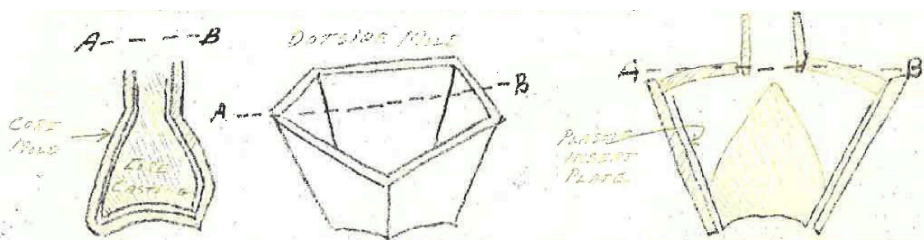

Figure 6. Kistiakowsky's hand drawn notes lens mold designs.[29]

## VI. COMPOSITION B, A NEW MOLECULE AND EXPLOSIVE FOR WORLD WAR II OFFERS A FAST EXPLOSIVE

Composition B, often referred to as Comp B, is a mixture of 59.5 wt% RDX, 39.5 wt% TNT, and 1.0 wt% wax desensitizer. It is one of the family of mixtures of RDX and TNT (molecular structures shown in Figure 7) known as cyclotols. Composition B has been a common melt-castable high explosive for a wide range of convention explosive ordnance, including artillery projectiles, rockets, land mines, hand grenades and various other munitions. It is worth noting that unlike TNT that had been around since 1863, RDX was not initially developed until the early days of World War II by Britain's Woolwich Arsenal.[30] It offers about 1.5 times the explosive energy of TNT per unit weight and about 2.0 times per unit volume, so for security reasons, Britain named it "Research Department Explosive", which continues on as "RDX". The British discovered that by adding TNT to a mixture with RDX, the resulting product was more stable during shipping, and they named it Composition B.

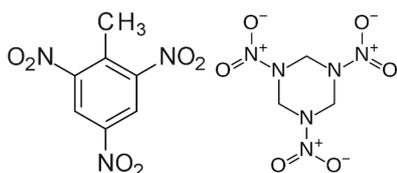

Figure 7. Molecular depiction of TNT (2,4,6-Trinitrotoluene), on the left, and RDX (1,3,5-Trinitro-1,3,5-triazinane), on the right.

Recognizing that Woolwich did not have the capacity to meet the Royal Air Force production needs for RDX, British leaders lobbied the U.S. to start production. However, in the early 1940s, the major U.S. explosive manufacturers—E. I. du Pont de Nemours & Company and Hercules—had several decades of experience of manufacturing TNT and were not interested in expanding into new explosives. U.S. Army Ordnance held the same viewpoint, wanting to continue using TNT. RDX had been tested by Picatinny Arsenal who regarded it as too expensive and too sensitive, and the Navy proposed to continue using ammonium picrate.[3]

Enter the NDRC and OSRD, who had visited The Royal Arsenal at Woolwich in the UK and thought new explosives were necessary for the war effort. James Conant, who recruited Kistiakowsky to the Manhattan Project, wished to involve academic research into this area and set up an experimental explosives research laboratory at the Bureau of Mines, Bruceton, Pennsylvania. NDRC research into new processes lead to major advances in RDX and Composition B production. The UK "Woolwich" method of manufacture of RDX was expensive, requiring 10 kg of strong nitric acid for every 1 kg of RDX[31], and there were no adequate sources of natural beeswax to desensitize the vast production of RDX. At least three laboratories at Cornell, Michigan, and Pennsylvania State universities[3] were instructed to develop better production methods for RDX. Werner Bachmann successfully developed the "combination process" by combining the Canadian process with direct nitration,[3, 32] which ultimately became known as the Bachmann process. A substitute stabilizer based on petroleum instead of beeswax was developed at the Bruceton Explosives Research Laboratory.[3]

Kistiakowsky indicated Composition B was chosen for the Gadget and Fat Man because one of his "associate organic chemists, primarily Dr. Bachmann of the University of Michigan, discovered the so-called Process B—Bachmann—of making RDX relatively economically" going back to his connections from his NDRC and OSRD days. Based on these advances, and being convinced of Composition B's superior qualities, the Army hired the Eastman Kodak Company of Rochester, New York, to manage production of RDX. Kodak, established a subsidiary, the Tennessee Eastman Company, to operate what is now known as the Holston Army Ammunition Plant. The site is presently the primary U.S. supplier of explosive materials for the Department of Defense (DOD) and Department of Energy (DOE)-National Nuclear Security Administration (NNSA), and is today the sole U.S. producer for RDX, HMX and 1,3,5-Triaminotrinitrobenzene (TATB), as well as formulated molding powder for PBX 9501 and PBX 9502. While Composition B is still widely used in conventional munitions, the DOD is slowly replacing Composition B with insensitive high explosive (IHE) formulations such as with IMX-101 (Insensitive Munitions Explosive) in U.S. military artillery shells and with IMX-104 in mortar rounds and hand grenades.[33] For planewave lenses to support Research and Development, in the mid-1980s LANL changed over to using pressed plastic bonded explosive PBX 9501 for the fast component.[28]

From Gibbs and Popolato,[34] Composition B is manufactured by melting the TNT in a steam-jacketed kettle equipped with a stirrer, brought to about 100°C. Water-wet RDX is slowly added. The mixture is heated and stirred until most of the water is evaporated, at which point the desensitizing wax is thoroughly mixed in. The material is typically cast into strips or chips to be later remelted and either directly cast into munition cases or into desired shapes. To achieve higher densities, a vacuum may be applied to the melted Composition B before to casting. Composition B has a theoretic density of 1.737 g/cm$^3$ and commonly achieves an open melt cast density or 1.68 to 1.70 g/cm$^3$ or a vacuum melt cast density of 1.715 to 1.720 g/cm$^3$. It melts around 79°C. The nominal detonation velocity is 8.018 mm/µs with a 29.22 GPa detonation pressure, from Deal.[35] The shock Hugoniot has a shock velocity $U$s = 2.71 + 1.86 $U$p for particle velocity values of 0 < $U$p < 1 mm/µs, from Coleburn and Liddiard.[36] The drop weight sensitivity impact height is 59 cm for a type 12 tool and 109 cm for a type 12B tool. Composition B has ultimate tensile strength of 0.9 to 1.0 MPa, ultimate compressive strength of 11.4 to 14.2 MPa, and compressive modulus of 14.5 to 18.6 GPa. However, at 50°C the ultimate compressive strength is reduced to 9.0 to 10.3 MPa and compressive modulus of 4.3 to 4.4 GPa. Significant work continues at LANL into characterizing Composition B, including understanding the effect of temperature on molten viscosity,[37] using x-ray scattering to probe the structural evolution of detonation carbon,[38] and measuring the effect of hot isostatically pressing on shock initiation.[39]

## VII. BARATOL, CALLING ON OLD FRIENDS FOR A SLOW TUNABLE EXPLOSIVE

The second (slow) type of explosive in the lenses was Baratol. Baratol is a melt castable high explosive mixture of barium nitrate (molecular structures shown in Figure 8) and TNT and actually refers the all ratios of those two components. During World War II the British developed Baratol formulations with roughly 20 wt% barium nitrate. This was part of a broad effort to develop explosives that could replace straight TNT ordnance fills because of the shortage of TNT.

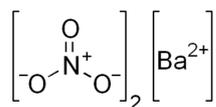

Figure 8. Barium dinitrate.

Similarly, the explosive Amatol consisting of a mechanics mixture of TNT and ammonium nitrate was developed during World War I and used extensively during World War II in mines and warheads. The U.S. used Baratols with slightly more barium nitrate to fill depth charges and other limited munitions. Needing an explosive of a very low detonation velocity to pair with the Composition B in the lenses for Fat Man, Kistiakowsky noted[40] that he "assigned the job to his friends at an Explosives Research Laboratory in Pittsburgh, the so-called Bruceton Laboratory, of which I was the chief until I went to Los Alamos." At the Pittsburgh laboratory, Dr. Duncan MacDougall and an associate developed the Baratol explosive containing up to 76 wt% of barium nitrate. Duncan MacDougall also ultimately made his way to Los Alamos, where he succeeded Kistiakowsky as the X Division Leader. In 1948 MacDougall lead the merger of X Division with the M (Implosion) Division ,which had been led by Darol Froman and the G (Gadget) Division which had been led Robert Bacher, to form GMX Division. MacDougall went on to lead GMX Division until 1970.

From Gibbs and Popolato,[34] Baratol is manufactured from finely ground barium nitrate added to molten TNT to form a castable slurry. About 0.1 wt% of nitrocellulose is added to the TNT before to adding the barium nitrate to reduce the slurry viscosity. Subsequent to the addition of the barium nitrate and just before applying a vacuum to the melt, 0.05 to 0.1 wt% of decylgallophenone or stearoxyacetic acid is added to prevent cracking. Before casting a vacuum is applied to the melt to remove dissolved and occluded gas to achieve a higher and more uniform density. Baratol produced with 76 wt% barium nitrate and 24% TNT has a theoretic density of 2.634 g/cm$^3$ and commonly achieves a vacuum cast density of 2.60 to 2.62 g/cm$^3$. It melts at a temperature of around 79 to 80°C. For a 76 wt% barium nitrate formulation with 2.61 g/cm$^3$. The nominal detonation velocity is 4.925 mm/µs with a 14 GPa detonation pressure, although the velocity is tunable by varying the percentage of barium nitrate. It is worth noting that during the Manhattan Project scientists needed to develop rate stick tests with an adequate booster to avoid anomalously low-velocity measurements from a point detonation and edge effects. They also found the refractive index between the fast and slow explosives could only be determined by experimental iteration, and the slow explosives velocity was observed to be faster in the lens configuration than measured from a rate stick test.[7] The shock Hugoniot is bilinear with a shock velocity $U$s = 2.40 + 1.66 $U$p for particle velocity values of 0 < $U$p < 0.75 mm/µs and $U$s = 1.50 + 2.16 $U$p for particle velocity values of 0.75 < $U$p < 1.25 mm/µs.[41] The drop weight sensitivity impact height is 110 cm for a type 12 tool and 140 cm for a type 12B tool. Baratol has an ultimate tensile strength of 2.6 to 3.1 MPa, an ultimate compressive strength of 39.3 to 55.8 MPa, and a compressive modulus of 10.3 to 13.8 GPa.

## VIII. WORKING AND LIVING WITH EXPLOSIVES, 1943-1945

Given an unparalleled opportunity to combine development of basic knowledge and the art of science to benefit the country and the allied war effort, many engineers and scientists, including George Kistiakowsky, spent many hours of every day and night working to develop and perfect an implosion design clearly intended to change the world as they knew it. Many scientists believed initially that it would be impossible to create a functional implosion atomic bomb. When it became clear that it could actually be built, they raced to develop it before Heisenberg and the Germans, and they hoped that the gadget they created would not just end the war, but end all "great" wars. Their spirits were high, and the enthusiasm for success had no bounds. The tempo, while unsustainable for an extended period of time, was extremely high, and generated many opportunities for ingenious improvisation and decision making. For balance, given limited vacation or travel opportunities, scientists and engineers made do with creative parties, hiking, and skiing in the nearby hills (Figure 9). Work was often merged with entertainment, and as the explosive's Division Leader, Kistiakowsky was a prime example of this. To quickly improve a ski run for the many Europeans working on the hill who loved skiing, Kistiakowsky creatively used surplus plastic explosives to cut down the trees. "If one builds a half necklace around the tree," he recalled, "then the explosion cuts it as if it were a chainsaw – and it's faster. A little noisier, though."[6]

Always a "hands-on" explosives expert, Kistiakowsky and his colleagues would X-ray the implosion shaped charges and identify flaws, such as air pockets. For the Trinity charges, hours before the test, he worked through the night using a dentist's drill to reach the voids to fill them with explosive. When asked about working with large quantities of explosives—during a time when the hazards of explosives were much less understood and there were not today's high explosive safety programs[42]—Kistiakowsky was quoted as saying, "I was completely confident. Besides, you don't worry about it. I mean if fifty pounds of explosive goes [off] in your lap, you have no worries."[40] Despite his pragmatic attitude towards explosives handling, he also wrote an excellent explosives safety manual in early 1945, detailing some of the safety concepts and issues that are still causing difficulties today, such as impact, burning, detonation, friction, and drop sensitivities.[43] In part thanks to his efforts, there were no explosive operations fatalities in Los Alamos during the Manhattan Project.

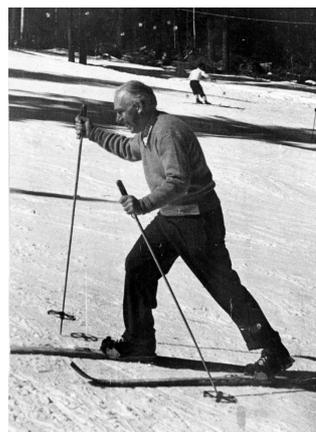
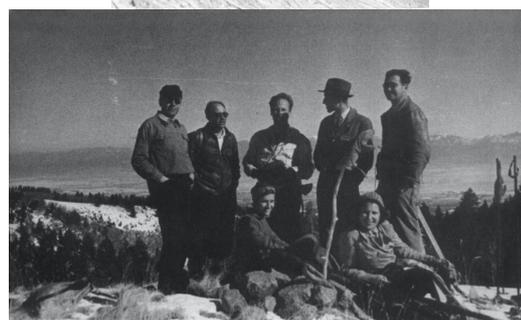

Figure 9. Niels Bohr skiing at Sawyer's Hill during the Manhattan Project (early 1945) (left). The photograph was classified until 1950 to protect the work performed at Los Alamos. The Los Alamos staff worked a six-day week, making Sundays the time for recreation. On a hike, from left to right, standing, Emilio Segré, Enrico Fermi, Hans Bethe, H. H. Staub, Victor Weisskopf; seated, Erika Staub, Elfriede Segré (right). (Source: AIP Emilio Segrè Visual Archives).

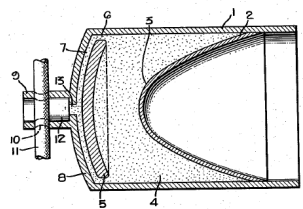
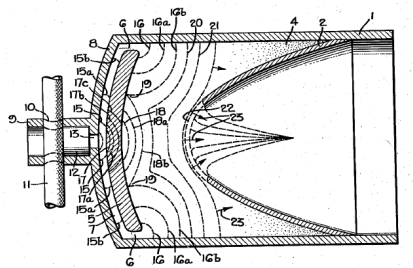

Figure 10. Shaped-charge patent from Haliburton Company[44] for perforation of well casing and subterranean formations surrounding oil, gas and water wells.

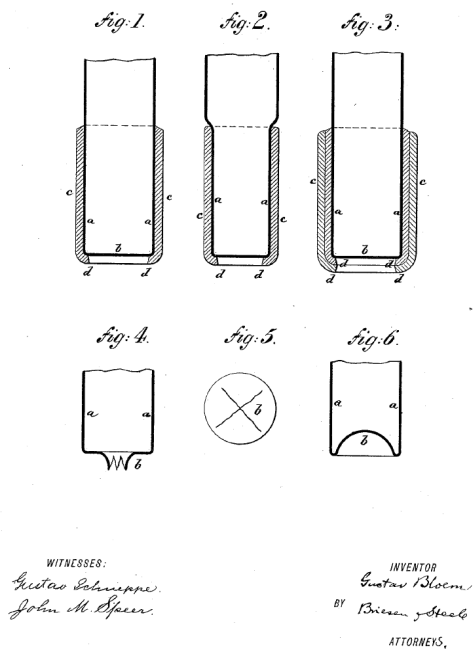

Figure 11. The 1886 patent showing detonator sleeves pressed inward hemispherically to improve performance.[45]

## IX. LASTING INFLUENCE OF DEVELOPMENT OF PRECISION EXPLOSIVES

While the implosion system for the Trinity Device was uniquely developed for the purpose of the atomic bomb, the concepts of engineered, high-precision explosives and directed-energy explosives has had broader impacts in the intervening 75 years.[46, 47] Going back to the early work of James Tuck on shaped charges[22] that were employed in anti-tank weapons by the British in World War II, high explosive lens concepts have been extensively employed in shaped charges for conventional military and civilian applications. A shaped charge is a specially designed class of explosive charges in which the geometry is selected to focus the effect of the energy coming from the detonation on a point or small area relative to the diameter of the charge (displayed in Figure 10). In bare charges with a void or material cut away at the surface, this focusing is known as the Munroe effect, named for Charles E. Munroe, a chemist at Naval Torpedo Station and War College at Newport, Rhode Island who discovered the effect in 1888.[48] While not named, the concept of a cutaway to compound the explosive energy had previously been employed, as an example for detonators in a 1886 patent,[45] show in Figure 11. Typically, a ductile metal liner is placed along the surface of the cutaway in the explosive, as either a simple right cone, a hemisphere, or a more complex tuned geometry, similar to the interface between the high- and low-velocity explosives in the high-explosive lens. When the explosive is detonated, the liner collapses along the center axis, with the converging material projecting a hypersonic jet along the axis of the charge capable of penetrating steel to depth multiple times the diameter of the charge. Although most shaped charges are a single explosive with the similarity to lenses coming through the timing of the detonation reaching the liner as it would reach the second explosive in a shaped charge, work has also been done with two explosives designs for shaped-charge jets.[49] While highly effective as a conventional munition[50] for penetrating heavily armored vehicles at close range, shaped charges are also an effective tool at breaching concrete. Extensive work has been performed by the DOD on development of shaped-charge jets and their interaction with targets, often employing DOE-NNSA modeling and simulation tools and experimental capabilities. Proton radiography (pRad) has been employed at Los Alamos to image shaped charge jet penetration into metal and ceramic armor materials,[51, 52] as shown in Figure 12. Because of the hazard from combatant or accidental impact of shaped charge jets into military explosive munitions, the North Atlantic Treaty Organization (NATO) has established a policy for introduction and assessment of Insensitive Munitions in Standardization Agreement (STANAG) 4439 to promote the development of safer munitions and increase interoperability with the established STANAG 4526, "Shaped Charge Jet Munition Test Procedures".[53] Explosive lenses are also employed to in conventional munitions to optimize the explosive transfer between booster and warhead.[54] In civilian applications shaped charges are commonly employed in mining and oil exploration as a way to quickly and effectively perforate geological materials, as illustrated in Figure 13.[55] While a single shaped charge can be placed to provide a more directed perforation when compared to a simple explosive charge, a perforating gun is frequently employed to provide effective flow communication between a cased wellbore and a productive reservoir. The perforating gun houses numerous shaped-charge pointing radially from the axis of the gun—typically 12 to 36 charges per meter each with 3 to 60 g of explosives—housed in a heavy walled tube. When fired most of the charge debris remains in the carrier, which is recovered from the well after firing. In addition to the axisymmetric shaped charges, linear shaped charges—with a plane of symmetry—create planar jets that can serve as a knife to cut through material. Linear shaped charges are commonly employed in demolition work (see Figure 14). Through precision application of linear shape charges, large structures can be felled in a controlled small foot print. As illustrated in the time lapse of the controlled demolition blasting of a chimney at the former Henninger brewery the multi-story tall chimney is controlled to fall almost entirely straight down. This type of precision demolition allows for large structures to be quickly and efficiently removed without affecting surrounding buildings. Higher precision applications of linear shaped charges include separating the stages of multistage rockets in flight. As illustrated in Figure 15,[56, 57] the modern SpaceX Crew Dragon capsule Flight Termination System (FTS) uses a linear-shaped charge assembly mounted onto the exterior case that houses the Falcon 9 rocket's motors and liquid fuel tanks.[58] The precision linear shaped charges allow for very rapid and clean separation of stages.

Planewave lenses themselves continue to be a valuable tool in research for creating a uniform shock loading to very high pressure. Baratol is now deemed a hazardous waste because of the barium content and numerous efforts have been undertaken to develop similar low-velocity explosives for plane-wave lenses, displayed in Figure 16, to support R&D work[26, 28] within the DOE-NNSA and DOD. Recent advances include the application of additive manufacturing (3-D printing) to simplify the production of plane-wave lenses.[59] Plane wave lenses themselves are frequently employed to either provide a direct high pressure impulsive Taylor wave load,[60-68] to launch a flier plate to achieve a high pressure square wave load,[69, 70] to achieve planer detonation of an explosive sample,[71, 52, 72] or to provide a uniform dynamic drive from explosive by-products[73, 74] for shock and detonation physics experiments.

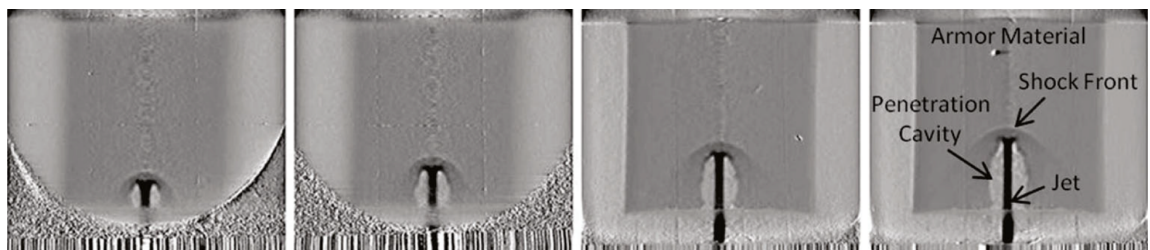

Figure 12. Movie of a metallic jet penetrating up into a cylindrically shaped armor material. Time increases from the left to right image.[52]

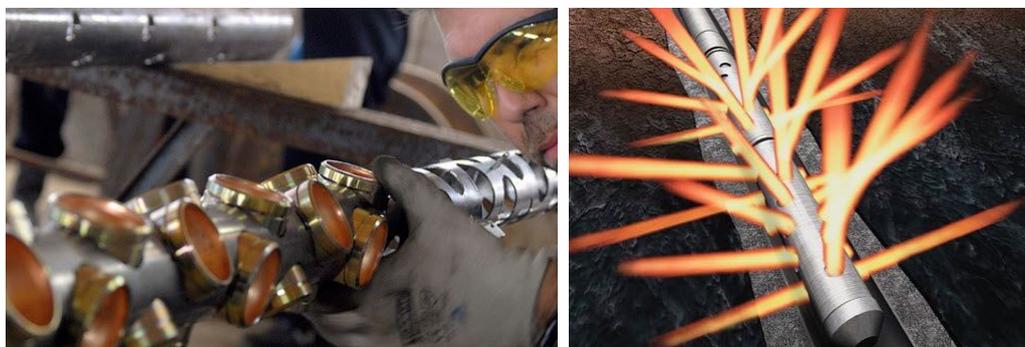

Figure 13. Perforating charges assembled into a cannon and an illustration of radial perforating of geological materials (Source: Petropedia).

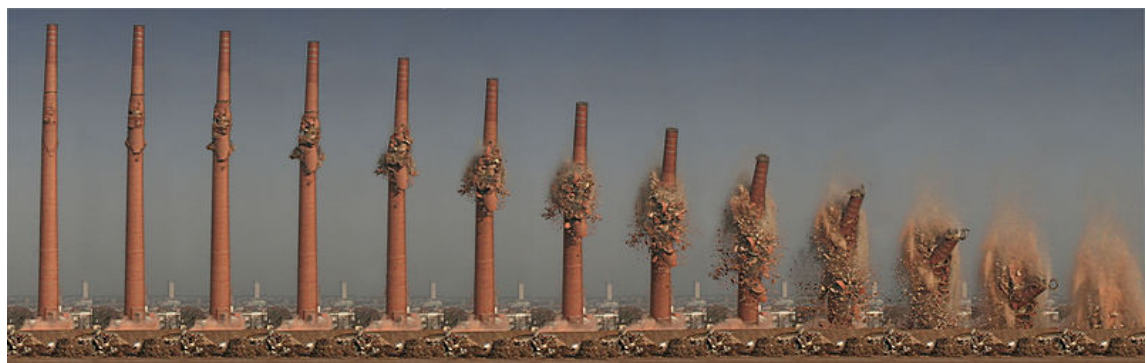

Figure 14. Time sequence of controlled demolition blasting of a chimney at the former Henninger brewery in Frankfurt am Main, Germany, Sachsenhausen (Source: Heptagon, Wikimedia Commons).

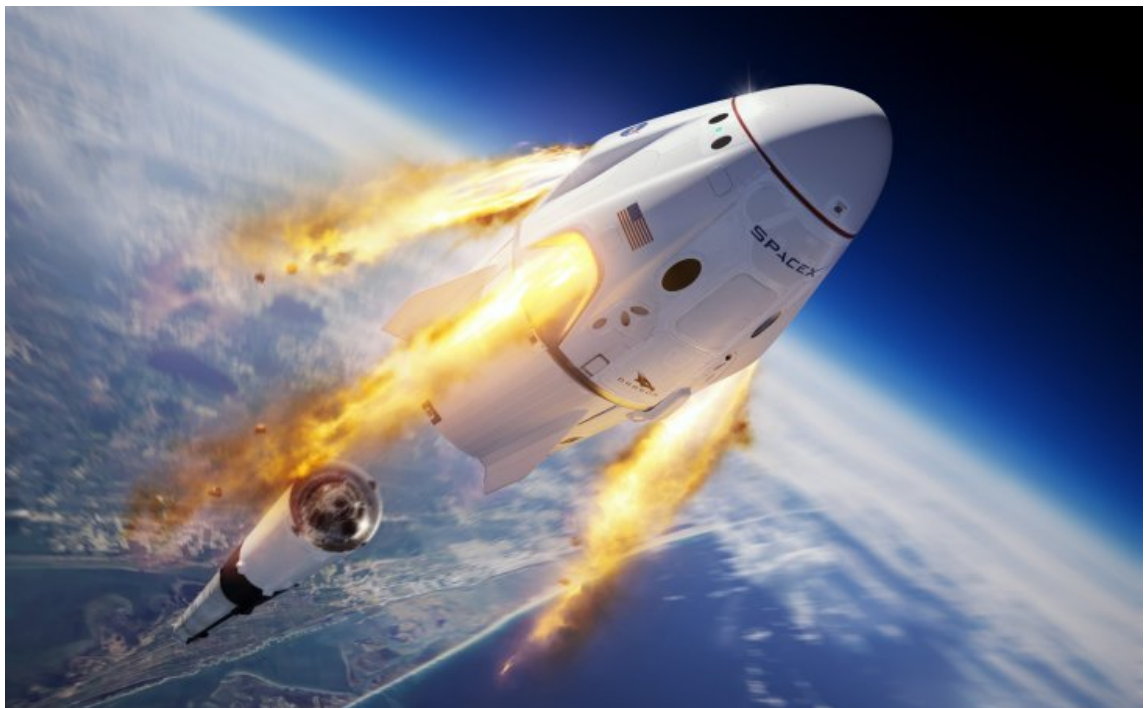

Figure 15. Illustration of SpaceX Crew Dragon capsule separating from the Falcon 9 rockets in flight (Source: SpaceX from the Associated Press). The Flight Termination System (FTS) on the inaugural Falcon 9 launch uses a linear-shaped charge assembly mounted onto the exterior case that houses the rocket's motors and liquid fuel tanks.[58]

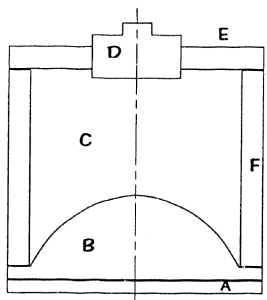

Figure 16. Schematic of Fritz plane wave lens.[26] A – acceptor explosive, B – plastic wave shaper, C – donor explosive, D – detonator, E – detonator support and lid for explosive, F – plastic cylinder to hold in explosive.

## X. CONCLUSION

The extraordinary effort, undertaken at Los Alamos in the mid-1940s to develop the Trinity implosion design, instigated new precision explosive design tools unlike any seen previously, including high-explosive synthesis and formulation, test and diagnostic facilities, shock dynamics theory, high-explosive system design engineering, and three-dimensional modeling and simulation. The high-explosive characterization and development process for precision implosion at Los Alamos, which began in 1944 under such dire circumstances, has persisted ever since, and brought with it many other constructive applications. It kindled a technical understanding of explosive behavior and design that allowed for a much broader application of precision high explosives in conventional munitions, demolition, mining and oil exploration, and space travel. These subsequent "constructive" applications, born from such destructive tools as explosives, owe their foundation to a highly-improbable secret project that began on a cottonwood-lined mesa in the northern high desert of New Mexico.

### Acknowledgments

This work was supported by the U.S. Department of Energy through the Los Alamos National Laboratory. Los Alamos National Laboratory is operated by Triad National Security, LLC, for the National Nuclear Security Administration of U.S. Department of Energy (Contract No. 89233218CNA000001). We would like to thank Mark Chadwick and Alan Carr for helpful discussions and Madeline Whitacre, Daniel Alcazar, and John Moore of the Los Alamos National Security National Security Research Center with finding and digitizing historical documents from Site Y of the Manhattan Project.

### ORCID

Eric N. Brown http://orcid.org/0000-0002-6812-7820